\documentclass[%
preprint,
superscriptaddress,
amsmath,amssymb,
aps,
]{revtex4-1}

\usepackage{graphicx}
\usepackage{dcolumn}
\usepackage{bm}

\usepackage[T1]{fontenc}       
\usepackage[utf8]{inputenc}
\usepackage{lmodern}%
\usepackage{hyperref}%
\usepackage{xcolor}%
\usepackage{placeins}
\usepackage{tabularx}
\usepackage{booktabs}
\usepackage{amsmath}

\usepackage[normalem]{ulem}

\newcommand{\beq}{\begin{equation}\begin{aligned}}
\newcommand{\eeq}{\end{aligned}\end{equation}}

\newcommand{\wse}{WSe\ensuremath{_2}}

\usepackage{setspace}

\DeclareFontFamily{U}{calligra}{}
\DeclareFontShape{U}{calligra}{m}{n}{<->callig15}{}
\newcommand{\calE}{{\!\!\text{\usefont{U}{calligra}{m}{n}E}\,\,}}

\newcommand{\dqmp}{Department of Quantum Matter Physics, University of Geneva, 24 Quai Ernest Ansermet, CH-1211 Geneva, Switzerland}
\newcommand{\gap}{Group of Applied Physics, University of Geneva, 24 Quai Ernest Ansermet, CH-1211 Geneva, Switzerland}
\newcommand{\ital}{Dipartimento di Scienze Fisiche, Informatiche e Matematiche,
University of Modena and Reggio Emilia, IT-41125 Modena, Italy}

\definecolor{linkcol}{rgb}{0,0,0.4}
\definecolor{citecol}{rgb}{0.5,0,0}

\hypersetup{
	bookmarksopen=true,
	pdftitle="Controlled band gap quenching in two-dimensional semiconductors",
	pdfauthor="Domaretskiy et al",
	pdfsubject="Controlled band gap quenching in two-dimentional semiconductors", 
	pdfstartview={FitH},    
	pdfmenubar=true, 
	pdfhighlight=/O, 
	colorlinks=true, 
	pdfpagemode=UseNone, 
	pdfpagelayout=SinglePage, 
	pdffitwindow=true, 
	linkcolor=linkcol, 
	citecolor=linkcol, 
	urlcolor=blue
}

\begin{document}
	
	
	\title{Quenching the band gap of 2D semiconductors with a perpendicular electric field}
	
	%
	%
	
	\author{Daniil Domaretskiy}
		\thanks{These authors have contributed equally}
	\affiliation{\dqmp}
	\affiliation{\gap}
	\author{Marc Philippi}
		\thanks{These authors have contributed equally}
	\affiliation{\dqmp}
	\affiliation{\gap}
	\author{Marco Gibertini}
	\affiliation{\dqmp}
	\affiliation{\ital}
	\author{Nicolas Ubrig}
	\affiliation{\dqmp}
	\affiliation{\gap}
	\author{Ignacio Gutiérrez-Lezama}
	\affiliation{\dqmp}
	\affiliation{\gap}
	\author{Alberto F. Morpurgo}
	\email{alberto.morpurgo@unige.ch}
	\affiliation{\dqmp}
	\affiliation{\gap}
	%
	
	
	\date{\today}
	
	
	\maketitle

	{\bfseries The electronic band structure of atomically thin semiconductors can be tuned by the application of a perpendicular electric field. The principle was demonstrated experimentally shortly after the discovery of graphene by opening a finite band gap in graphene bilayers\cite{oostinga_gate-induced_2008,zhang_direct_2009,mak_observation_2009}, which naturally are zero-gap semiconductors. So far, however, the same principle could not be employed to control a broader class of materials, because the required electric fields are beyond reach in current devices. To overcome this limitation, we have realized double ionic gated transistors  that enable the application of very large electric fields. Using these devices, we show that the band gap of few-layer semiconducting transition metal dichalcogenides can be continuously suppressed from 1.5 eV to zero. Our results illustrate an unprecedented level of control of the band structures of 2D semiconductors, which is important for future research and applications.}
	
An electric field applied perpendicular to the surface of a bulk semiconductor is screened over a finite length, leaving the material interior unaffected. In atomically thin semiconductors \cite{novoselov_two-dimensional_2005-1}, however, the small thickness prevents efficient screening, so that a perpendicular electric field uniformly influences the entire system, modifying its band structure\cite{ramasubramaniam_tunable_2011-1,drummond_electrically_2012,chu_electrically_2015,kim_observation_2015,dai_bandstructure_2015,deng_efficient_2017,overweg_electrostatically_2018}. Indeed, a zero-gap semiconductor such as bilayer graphene can be turned into a gapped insulator using double-gated transistors to apply a perpendicular electric field\cite{oostinga_gate-induced_2008,zhang_direct_2009,mak_observation_2009}. Despite representing a true conceptual breakthrough, continuous control of the band structure in transistors has not found widespread use, because the limited maximum electric field that can be applied in common devices does not allow significant changes to be induced in most 2D materials. 
	
\begin{figure}[h]
		\centering
		\includegraphics[width=\linewidth]{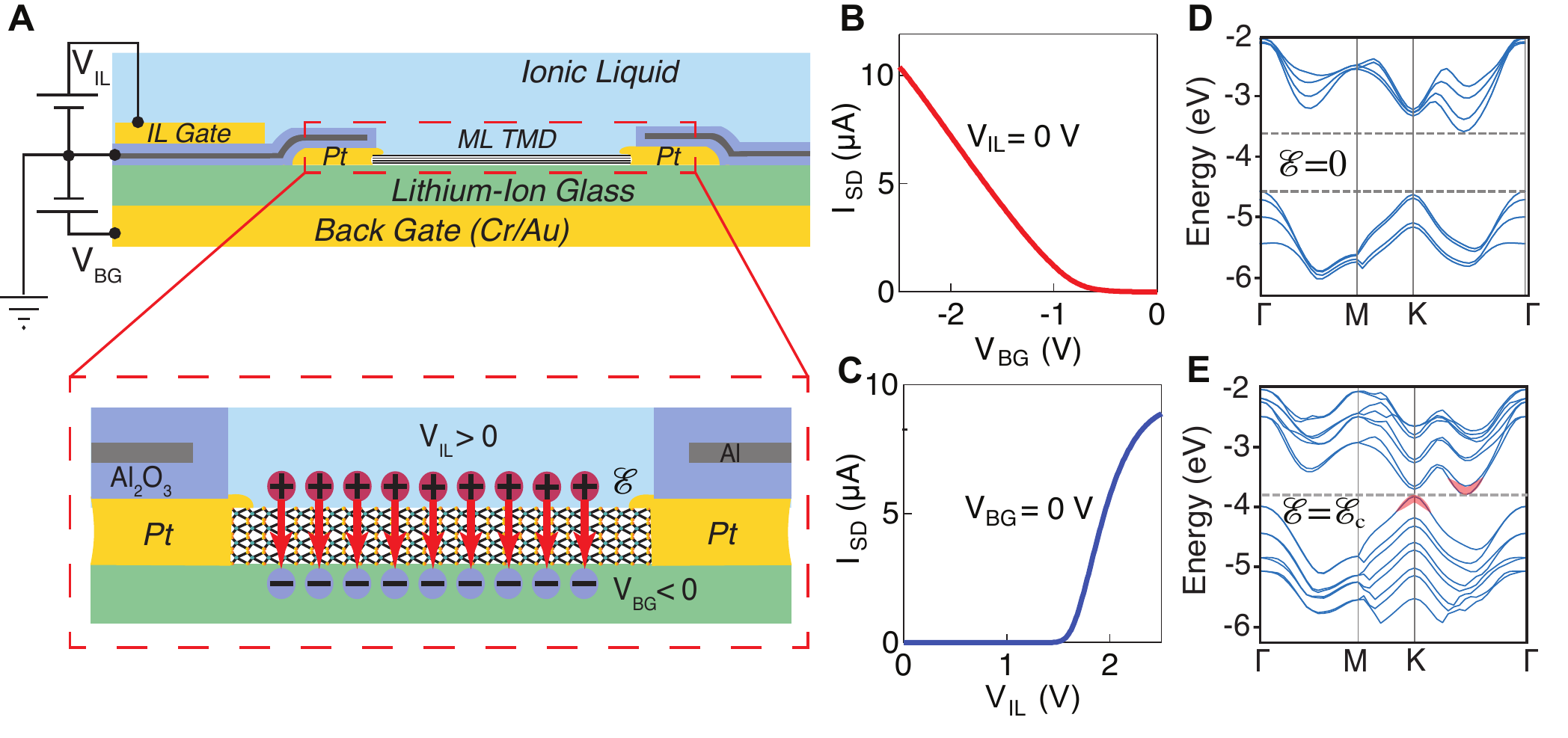}
		\caption{{\bfseries Double ionic gated field-effect transistors.} {\bfseries A} (top panel) Schematic cross-section of a multilayer WSe$_2$ transistor equipped with top (ionic liquid) and bottom (Li-ion conductive glass ceramic) electrolyte gates. Also shown are the Platinum contacts to the TMD multilayer and a Al$_2$O$_3$/Al/Al$_2$O$_3$ trilayer to decouple electrostatically the top and bottom electrolytes. (bottom panel) Zoom in on the device channel area (not to scale). When the two gates are biased with opposite polarity the accumulated charges compensate, and a uniform perpendicular electric field $\calE$ is established across the TMD (represented by the red arrows in the scheme). {\bfseries B} Source-drain current $I_{SD}$ measured for negative voltages $V_{BG}$ applied to the Li-ion glass gate (with the ionic liquid gate grounded, $V_{IL}$= 0 V), resulting in the accumulation of holes in WSe$_2$. {\bfseries C} $I_{SD}$ measured as  a function of $V_{IL} > 0$ V for $V_{BG}$ = 0 V, to cause electron accumulation in the transistor (the applied source-drain voltage is $V_{SD} =0.1$  V). Note that the application of positive $V_{IL}$ and negative  $V_{BG}$ < 0 V causes the Li ions to be pulled away from the TMD, ensuring that intercalation does not take place.  {\bfseries D} and {\bfseries E}  represent the band structure of 4L WSe$_2$ computed within density-functional theory --respectively for  zero perpendicular electric $\calE =0$ V/nm, and at the critical field $\calE =\calE_{c}$-- and show that quenching of the gap with an electric field  is expected theoretically  (see Supplementary Note S7 for details; in ({\bfseries E}), the conduction and valence band edge  at $\calE_{c}$  are  denoted by  red shaded lines). 
		}
		\label{fig:01}
\end{figure}

Here we demonstrate the ability to fully quench band gaps as large as 1.5 eV using a new type of transistors based on double ionic gates. Ionic gating relies on electrolytes to transfer the potential from a metallic gate to the surface of a semiconductor\cite{fujimoto_electric-double-layer_2013,schmidt_characterization_2016,bisri_endeavor_2017}. In our devices, two electrolytes in contact with two independent gate electrodes are coupled to a same atomically thin semiconductor (see Fig. \ref{fig:01}\textbf{A}). The top electrolyte is a commonly used ionic liquid\cite{fujimoto_electric-double-layer_2013,schmidt_characterization_2016,bisri_endeavor_2017,zhang_band_2019,gutierrez-lezama_ionic_2021}, and the bottom one is a Li-ion glass ceramic substrate\cite{philippi_lithium-ion_2018,alam_lithium-ion_2020}. The atomically thin semiconductor is also connected to two metal electrodes functioning as source and drain contacts, and surrounded by a ground plane (an Aluminum film sandwiched between two $\mathrm{Al_2O_3}$ layers) to eliminate any direct electrostatic coupling of the top and bottom electrolyte (See Supplementary Note S1 for details on device fabrication and S3 for the decoupling of the gates). The use of single ionic gate devices to accumulate charge densities unattainable with conventional gating is by now established\cite{ye_superconducting_2012,wang_hopping_2012,fujimoto_electric-double-layer_2013,lu_evidence_2015,costanzo_gate-induced_2016,bisri_endeavor_2017, leighton_electrolyte-based_2019-1}. However, double ionic gate devices to control independently accumulated charge and electric field have never been reported earlier. 

\begin{figure}
		\centering
		\includegraphics[width=\linewidth]{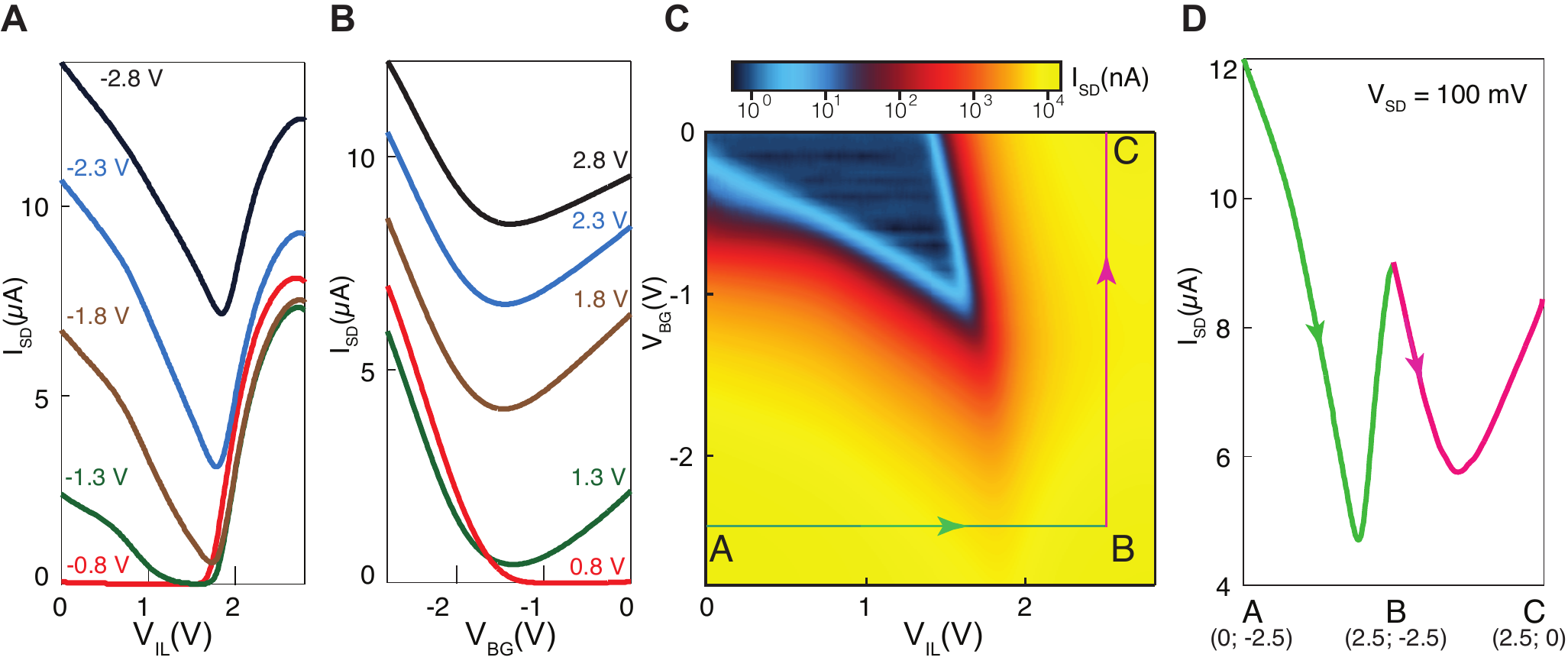}
		\caption{{\bfseries Electrical characteristics of a double-gated  4L-WSe$_2$ transistor.} \textbf{A} Source-drain current $I_{SD}$ as a function of ionic liquid gate voltage $V_{IL}$, for different negative values of back gate voltage $V_{BG}$. The curves evolve from exhibiting  textbook transistor behavior at small negative $V_{BG}$ (see also Fig. 1\textbf{B}), to not showing any sizable current suppression at large  negative $V_{BG}$.  \textbf{B} Same as \textbf{A}, with the source-drain current  $I_{SD}$ measured as a function of $V_{BG}$ for different positive values of $V_{IL}$; the evolution of the transistor curves is fully analogous to the one shown in \textbf{A}.  \textbf{C} Color plot of $I_{SD}$ (in logarithmic scale) as a function of $V_{BG}$ and $V_{IL}$.
		Note that the simultaneous application of a large negative $V_{BG}$ and of an equally large positive $V_{IL}$ causes the current in the transistor to increase by 4-to-5 orders of magnitude, despite leaving the potential of the transistor channel ($V^* = V_{IL}+V_{BG}$) unchanged. \textbf{D} Evolution of $I_{SD}$ along the A-B-C contour illustrated in panel (\textbf{D}) (the coordinates  of  A, B, and C in the ($V_{IL}$, $V_{BG}$) plane are indicated at the bottom). Transport in the transistor   is mediated by holes at  A and by electrons at C: finding that nowhere  the current is fully suppressed  implies that on part of the A-B-C contour the valence and conduction band of WSe$_2$ must overlap (i.e., the band gap closes). In all measurements  $V_{SD} = 0.1$ V.}
		\label{fig:02}
\end{figure}

When grounding one of the gates and sweeping the voltage applied to the other, these double gate devices function as conventional transistors, as illustrated by the transfer curves ($I_{SD}$ measured as a function of gate voltage) recorded on a representative \wse~tetra-layer (4L) device (see Fig. \ref{fig:01}\textbf{B} and \ref{fig:01}\textbf{C}; for additional characterization measurements, see Supplementary Note S2). Simultaneously applying a positive voltage $V_{IL}$ to the ionic liquid gate and a negative voltage $V_{BG}$ to the back gate also enables a perpendicular electric field to be established with no net charge accumulated on the semiconductor. To investigate the effect of the perpendicular electric field on the band structure, we look at how the transfer curves $I_{SD}$-vs-$V_{IL}$ of our device evolve for increasing negative $V_{BG}$ values (Fig. \ref{fig:02}\textbf{A}). 

For $V_{BG}$ = -0.8 V, the device transfer curve is qualitatively identical to that measured for $V_{BG}$ = 0 V (Fig. \ref{fig:01}\textbf{C}): the current $I_{SD}$ increases as $V_{IL}$ is swept past the threshold for electron accumulation ($V_{IL}$ = 1.7 V), and no current flows for $V_{IL}$ < 1.7~V, when the chemical potential is in the gap. At $V_{BG}$  $\sim$ -1 V, however, the transfer curve exhibits clear qualitative differences, as the current flows even for $V_{IL}$  well below 1.7 V. As $V_{BG}$ is further increased to more negative values, the source-drain current $I_{SD}$ remains large for all values of $V_{IL}$, without ever vanishing (with a square resistance $R_{sq} \sim 10~k\Omega$ for all $V_{IL}$). Analogous considerations hold true when looking at the evolution of the $I_{SD}$-vs-$V_{BG}$ transfer curve upon applying a positive voltage $V_{IL}$ to the ionic liquid gate. 

The complete evolution of $I_{SD}$ as a function of $V_{IL}$ and $V_{BG}$ is illustrated by the color plot in Fig. \ref{fig:02}\textbf{C}. The observed behavior is not the one expected if the only effect of the gate voltages was to affect the electrostatic potential in the transistor channel, i.e., the mechanism that determines the operation of a conventional transistor. In that case, both $V_{IL}$ and $V_{BG}$ would just shift the energy of the 4L-\wse~  bands, and a change in the voltage applied to one of the gates would only cause a rigid shift in the transfer curve measured as a function of the voltage applied to the other gate. The data, however, do not show a rigid shift. That is because the concomitant application of large positive $V_{IL}$ and large negative $V_{BG}$ also generates a perpendicular electric field that quenches the band gap, in agreement with previous theoretical calculations\cite{ramasubramaniam_tunable_2011-1,dai_bandstructure_2015} and our own (see Fig. 1\textbf{D} and 1\textbf{E} and Supplementary Note S7).

To understand, we look at the evolution of transport as $V_{IL}$ and $V_{BG}$ are varied continuously along the contour outlined by the colored line in Fig. \ref{fig:02}\textbf{C}. At point A, $V_{IL}~=~0 V$ and $V_{BG}$ =~-2.5 V. The negative potential $V_{BG}$ results in the accumulation of holes (see Fig. \ref{fig:01}\textbf{B}), and sets the chemical potential in the \wse~ valence band (the square resistance is $R_{sq}$ $\sim$ 8 k$\Omega$, Fig. \ref{fig:02}\textbf{B}). As $V_{IL}$ is increased from 0 to 2.5 V at fixed $V_{BG}$ =~-2.5~V, we travel from points A to B (green line). Since the capacitance of the ionic liquid and of the Li-ion glass gates are the same to a very good approximation, the electrostatic potential applied to the channel is proportional to $V^\star$ = $V_{IL}$+$V_{BG}$, and goes back to zero at B. Nevertheless, the current remains large. If then $V_{BG}$ is decreased from -2.5 to 0~V at fixed $V_{IL}$ = 2.5~V we move from B to C (purple line), where transport is mediated by electrons accumulated by the large positive voltage $V_{IL}$ (see Fig. \ref{fig:01}\textbf{C}). Transport therefore evolves from being mediated by holes at point A (with the chemical potential in the valence band), to being mediated by electrons at point C (with the chemical potential in the conduction band), without ever passing through a highly insulating state (see Fig. \ref{fig:02}\textbf{D}). This is possible only if the gap closes and the conduction and valence band overlap in some part of the contour, with electrons and holes coexisting in the transistor channel. Indeed, this happens in the neighborhood of B, where the electric field perpendicular to the 4L-\wse –proportional to $E^\star$ = ($V_{IL}$ - $V_{BG}$)– is maximum. 

\begin{figure}
		\centering
		\includegraphics[width=\linewidth]{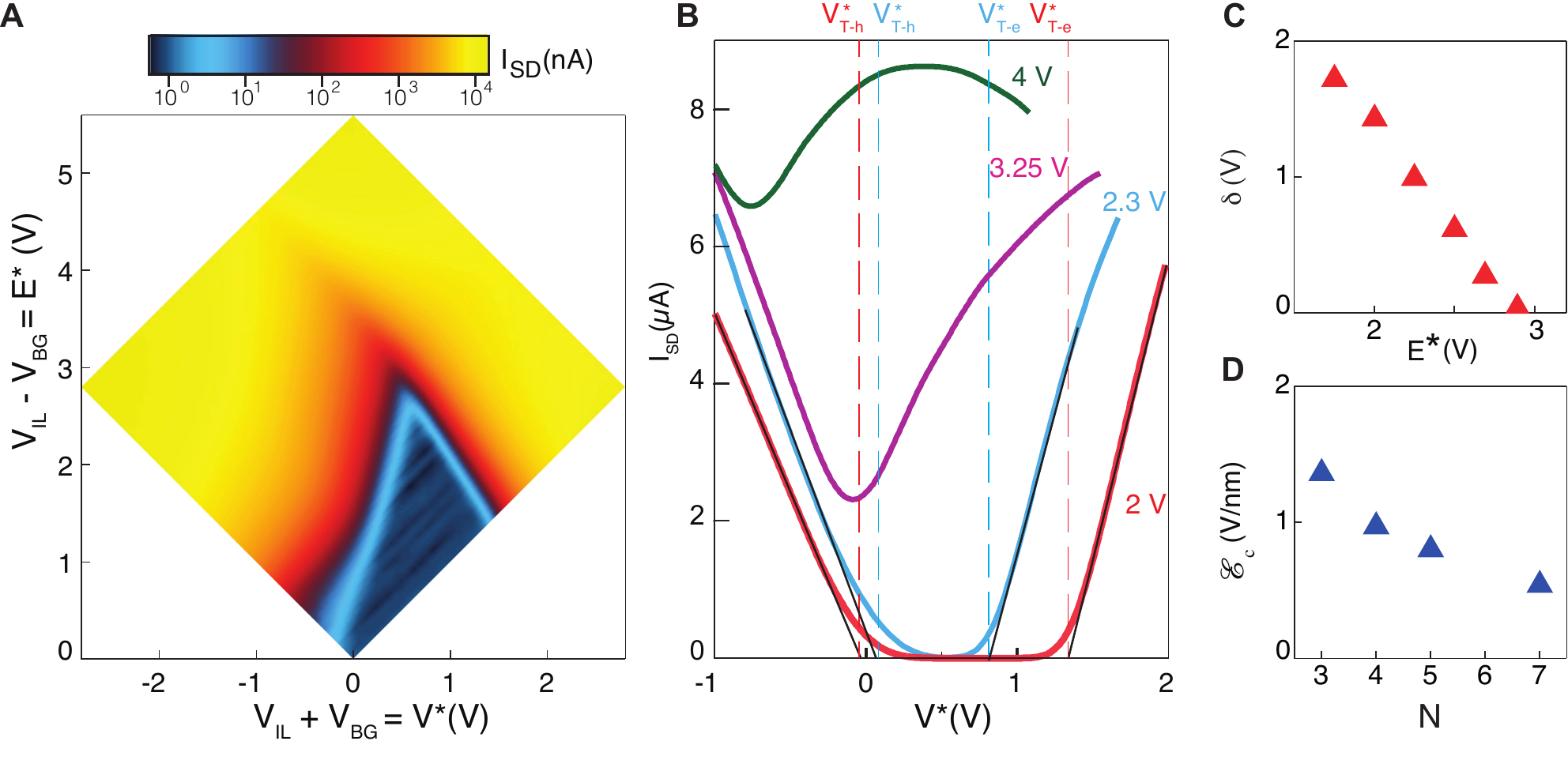}
		\caption{{\bfseries Band gap evolution as a function of electric field.} {\bfseries A} Colour plot of the source-drain current $I_{SD}$ (in logarithmic scale) as a function of  $V^*=V_{IL}+V_{BG}$,  and $E^*=V_{IL}-V_{BG}$ (respectively proportional to the electrostatic potential in  the transistor channel and to the  electric field perpendicular to the  4L-WSe$_2$ crystal). The width of the $V^*$ interval over  which the current $I_{SD}$ is suppressed decreases monotonically upon increasing $E^*$. {\bfseries B} For a quantitative analysis, we look at horizontal cuts of the color plot in ({\bfseries A}),  $I_{SD}$-vs-$V^*$ at fixed $E^*$ (values indicated in the figure). The curves measured at $E^* = 2.0$ and $2.3$ V exhibit a complete suppression of $I_{SD}$, and the black thin lines illustrate how we determine the threshold voltage for electron ($V^*_{T-e}$) and hole ($V^*_{T-h}$) conduction (by extrapolating $I_{SD}$ to zero for positive and negative values of $V^*$; the position of the threshold voltages are marked by the vertical dashed lines). {\bfseries C} We plot  $\delta = V^*_{T-e}-V^*_{T-h}$ as a function of $E^*$, and find the value  $E^*_c$ for which $\delta=0$ V ( $E^\star_c$ = 2.9 V in our 4L-WSe$_2$ device) to determine the condition at which the gap closes. {\bfseries D} The voltage difference $E^*_c$ measured for WSe$_2$ crystals with N~=~3,4,5,7 layers is converted into an electric field $\calE$, by modeling our device as a stack of three dielectrics, corresponding to the double layers of the two electrolyte gates and the WSe$_2$ multilayer (the very large, known capacitance of the electrical double layers makes the procedure rather insensitive to the value of the dielectric constant of the TMD; see Supplementary Note S5 for details). Experimental data on 3L, 5L and 7L WSe$_2$ devices is presented in Supplementary Note S6. 
		}
		\label{fig:03}
\end{figure}

For a quantitative analysis, we plot the current $I_{SD}$ as a function of $V^\star = V_{IL}$+$V_{BG}$ and $E^\star = V_{IL}$ - $V_{BG}$ (Fig. \ref{fig:03}\textbf{A}). Fig. \ref{fig:03}\textbf{B} shows $I_{SD}$-vs-$V^\star$ curves measured at fixed values of $E^*$. At small $E^\star$, $I_{SD}$ is finite for sufficiently large negative and positive $V^\star$  –with current mediated by holes and electrons respectively– and vanishes over an extended $V^\star$ interval, as the chemical potential in 4L-\wse~ is swept across the gap. This interval shrinks as $E^\star$ increases, and for $E^\star$> $E^\star_c$, no highly resistive state is observed, indicating that under these conditions no gap is present. To determine $E^\star_c$, we extract the threshold voltages for electron and hole conduction ($V^\star_{T-e}$ and $V^\star_{T-h}$), plot the difference $\delta =V^\star_{T-e} - V^\star_{T-h}$ as a function of $E^\star$, and look at when $\delta$ vanishes (Fig. \ref{fig:03}\textbf{C}). The corresponding electric field is then approximately given by $E^\star_c/t_{TMD}~\simeq~1.1$ V/nm, where $t_{TMD}$ = 2.6 nm is the thickness of 4L-\wse. This is a slight overestimate, because part of the voltage drops across the screening length of the two electrolytes. However, as we discuss in detail in Supplementary Note S5,  the very large capacitance of the gate electrolytes ($C\approx 50~\mu F/cm^2$ \cite{yuan_high-density_2009,fujimoto_electric-double-layer_2013,schmidt_characterization_2016,philippi_lithium-ion_2018,zhang_band_2019}) ensures that the voltage drop across the electrolytes is small,  only approximately 10 \%, from which we conclude that the electric field $\calE_c$ needed to quench the gap of 4L \wse~is between  0.9 and 1.0 V/nm (note that this is the actual electric field, and not the displacement field, which is frequently referred to in the literature when analyzing experiments on double-gated transistors). We have performed similar experiments also on 3L, 5L and 7L \wse~ and succeeded in closing the band gap in all cases (see Supplementary Note S6 for data measured on devices of different thickness). The required electric $\calE_c$ fields are plotted in Fig. \ref{fig:03}\textbf{D}. 

\begin{figure}
		\centering
		\includegraphics[width=\linewidth]{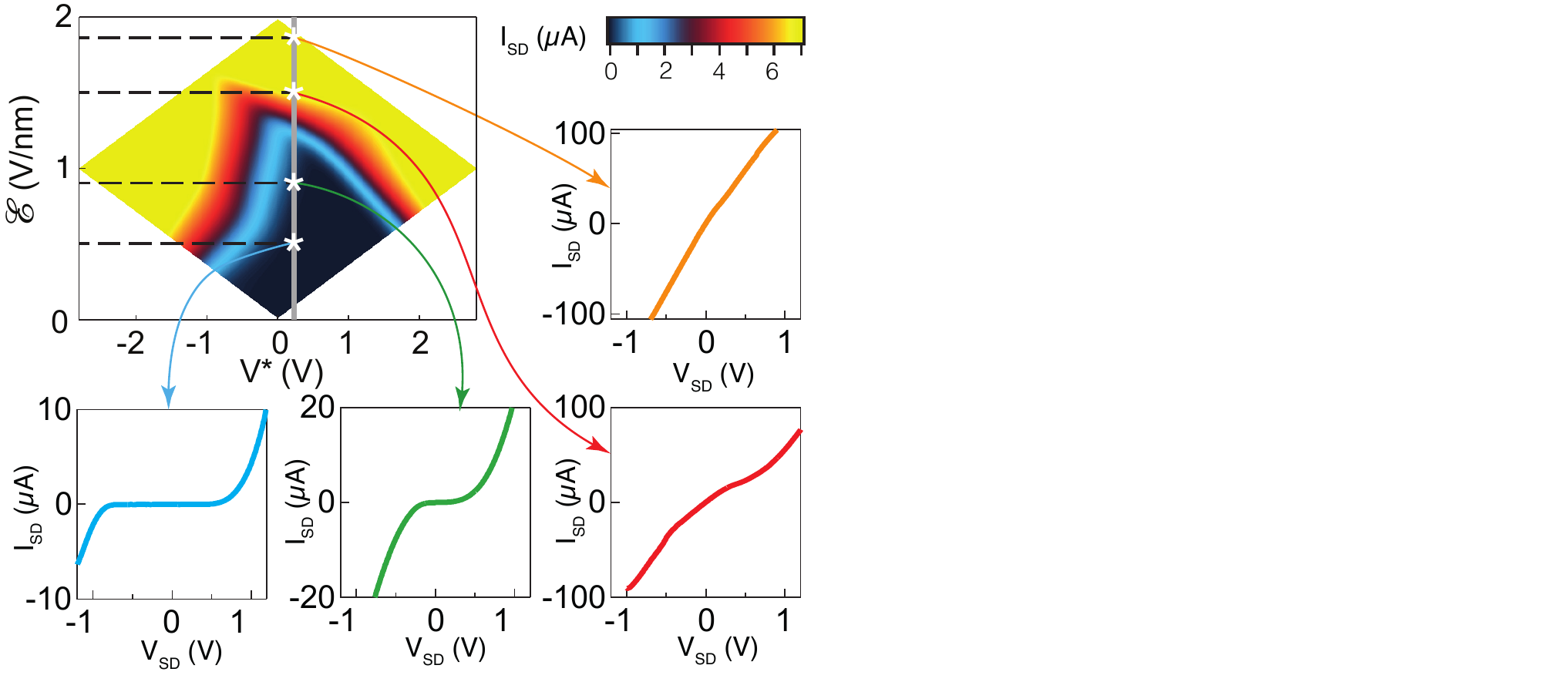}
		\caption{ {\bfseries Evolution of the $I_{SD}$-vs-$V_{SD}$ curves.} The top left panel is a linear-scale color plot of $I_{SD}$ as a function of $V^*=V_{IL}+V_{BG}$ and of perpendicular electric field $\calE$. The other surrounding panels --with curves of different color-- show the $I_{SD}$-vs-$V_{SD}$ characteristics of our transistor measured at $V^*$ = 0.2 V for different values of $\calE$ (indicated by the white starts in the color plot). At $\calE = 0.5$ V/nm (blue curve) a very strong non-linearity with a large suppression of $I_{SD}$ at small $V_{SD}$ is clearly visible, as expected (because $V^*$ = 0.2 V corresponds to having the chemical potential inside the band gap of 4L-WSe$_2$). Upon increasing the perpendicular electric field, the non-linearity in the $I_{SD}$-vs-$V_{SD}$ characteristics becomes less pronounced (green curve, $\calE = 0.87$ V/nm). Increasing the perpendicular electric field further fully quenches the gap, causing the suppression of $I_{SD}$ at low $V_{SD}$ to become negligible (red curve, $\calE = 1.5$ V/nm), and eventually resulting in a virtually perfectly linear $I_{SD}$-vs-$V_{SD}$ curve  (orange curve, $\calE = 1.9$ V/nm).    }
		\label{fig:04}
\end{figure}

The evolution of the $I_{SD}$-vs-$V_{SD}$ curves upon increasing the perpendicular electric field confirms that the gap closes (see Fig. \ref{fig:04}). At $E^\star$ = 0, a pronounced suppression of $I_{SD}$ over a broad $V_{SD}$ range is observed when the 4L-\wse~device is biased at $V^\star$ = $V_{IL}$+$V_{BG}$~=~0.2~V, corresponding –for $V_{SD}$ = 0 V– to setting the chemical potential well inside the gap. Upon increasing $E^\star$ (with $V^\star$ fixed), the $I_{SD}$ suppression becomes less pronounced and eventually the $I_{SD}$-vs-$V_{SD}$ curve becomes perfectly linear. That is what happens after the gap has closed, when the overlap of valence and conduction band is sufficiently large.

Within the precision of both theory and experiments, the value of electric field needed to quench the gap of  few-layer \wse~ is in agreement  with expectations (See Supplementary Note S7). In qualitative terms, the gap closes when the electrostatic potential is sufficient to lift the energy of the valence band edge at one crystal surface above the conduction band edge at the opposite surface, so that the conduction and valence band overlap, and the system becomes gapless. This conclusion holds true for all semiconductors, but the nature of the gapless state depends on details. For sufficiently thin crystals –such as 4L-\wse– the electron and hole wavefunctions at opposite surfaces are expected to overlap in space and hybridize. In this regime, the appearance of topological states has been predicted in semiconducting transition metal dichalcogenides\cite{qian_quantum_2014,tong_topological_2017,zhu_gate_2019}. For thicker crystals hybridization eventually becomes negligible, and the electron and hole accumulation layers on opposite surfaces only couple through Coulomb interaction. Closing the gap of atomically thin semiconductors with a perpendicular electric field therefore leads to the formation of electron-hole systems that have not been realized earlier.

Tuning the gap continuously –and suppressing it altogether– discloses the possibility to control virtually all aspects of the physics of a semiconductor, including its emission and absorption spectrum, as well as other optical and transport properties. As such, the electrostatic control of the band gap opens new routes for the realization of technologically relevant photonic devices. Examples are broadly tunable light sources\cite{jauregui_electrical_2019}, possibly covering energy ranges in which achieving light emission is traditionally difficult (e.g., in the very far infrared), or optical modulators. We therefore anticipate that the results presented here will attract  interest in the fields of optoelectronics and optical communications.

\section*{Data availability}
The data that support the findings of this study are available from the corresponding author without any restriction.

\section*{Author contributions}
D.D., M.P., N.U., and I.G.L. fabricated the devices and carried out the experiments. D.D, M.P., I.G.L, and A.F.M. analysed the experimental data. M.G. performed first-principles simulations. I.G.L. supervised the experimental work and A.F.M. conceived and directed the research. All authors participated in preparing the manuscript. 

\section*{acknowledgements}
The authors thank A.~Ferreira for technical support. A.F.M. acknowledges financial support from the Swiss National Science Foundation (Division II) and from the EU Graphene Flagship project. M.G.\ acknowledges support from the Italian Ministry for University and Research through the Levi-Montalcini program.

\section*{Competing interests}
The authors declare no competing interests.

\newpage
\newpage

\newpage
\newpage

\clearpage
\bibliography{MainB}
\bibliographystyle{naturemag}

\end{document}